\documentclass[twoside,twocolumn,9pt]{article}
\usepackage[super,sort&compress,comma]{natbib} 
\usepackage[left=1.5cm, right=1.5cm, top=1.785cm, bottom=2.0cm]{geometry}
\usepackage{times,mathptmx}
\usepackage{sectsty}
\usepackage{graphicx} 
\usepackage{lastpage}
\usepackage{float}
\usepackage{fancyhdr}
\usepackage{fnpos}
\usepackage[english]{babel}
\usepackage[usenames,dvipsnames]{xcolor}
\usepackage{setspace}
\usepackage[compact]{titlesec}

\definecolor{cream}{RGB}{222,217,201}

\begin{document}

\pagestyle{fancy}


\makeFNbottom
\makeatletter
\renewcommand\LARGE{\@setfontsize\LARGE{15pt}{17}}
\renewcommand\Large{\@setfontsize\Large{12pt}{14}}
\renewcommand\large{\@setfontsize\large{10pt}{12}}
\renewcommand\footnotesize{\@setfontsize\footnotesize{7pt}{10}}
\makeatother

\renewcommand{\thefootnote}{\fnsymbol{footnote}}
\renewcommand\footnoterule{\vspace*{1pt}%
\color{cream}\hrule width 3.5in height 0.4pt \color{black}\vspace*{5pt}} 
\setcounter{secnumdepth}{5}

\makeatletter 
\renewcommand\@biblabel[1]{#1}            
\renewcommand\@makefntext[1]%
{\noindent\makebox[0pt][r]{\@thefnmark\,}#1}
\makeatother 
\renewcommand{\figurename}{\small{Fig.}~}
\sectionfont{\sffamily\Large}
\subsectionfont{\normalsize}
\subsubsectionfont{\bf}
\setstretch{1.125} 
\setlength{\skip\footins}{0.8cm}
\setlength{\footnotesep}{0.25cm}
\setlength{\jot}{10pt}
\titlespacing*{\section}{0pt}{4pt}{4pt}
\titlespacing*{\subsection}{0pt}{15pt}{1pt}

\fancyfoot{}
\fancyfoot[RO]{\footnotesize{\sffamily{\thepage\ / \pageref{LastPage} }}}
\fancyfoot[LE]{\footnotesize{\sffamily{\thepage\ / \pageref{LastPage} }}}
\fancyhead{}
\renewcommand{\headrulewidth}{0pt} 
\renewcommand{\footrulewidth}{0pt}
\setlength{\arrayrulewidth}{1pt}
\setlength{\columnsep}{6.5mm}
\setlength\bibsep{1pt}

%
%
%
%

\twocolumn[
  \begin{@twocolumnfalse}
\vspace{3cm}
\sffamily

\noindent\LARGE{\textbf{Self-assembly of a space-tessellating structure in the binary system of hard tetrahedra and octahedra$^\dag$}} \\
\vspace{0.3cm} 

  \noindent\large{A.\ T.\ Cadotte,\textit{$^{a}$}$^\ddag$, J.\ Dshemuchadse,\textit{$^{b}$}$^\ddag$ P.\ F.\ Damasceno,\textit{$^{a}$}$^\mathsection$ R.\ S.\ Newman,\textit{$^{b}$} and S.\ C.\ Glotzer$^{\ast}$\textit{$^{a,b,c,d}$}} \\

 \noindent\normalsize{

We report the formation of a binary crystal of hard polyhedra due solely to entropic forces.
Although the alternating arrangement of octahedra and tetrahedra is a known space-tessellation, it had not previously been observed in self-assembly simulations.
Both known one-component phases -- the dodecagonal quasicrystal of tetrahedra and the densest-packing of octahedra in the Minkowski lattice -- are found to coexist with the binary phase.
No additional crystalline phases were observed.
} \\


 \end{@twocolumnfalse} \vspace{0.6cm}

  ]

\renewcommand*\rmdefault{bch}\normalfont\upshape
\rmfamily
\section*{}
\vspace{-1cm}


\footnotetext{\textit{$^{a}$~Applied Physics Program, University of Michigan, Ann Arbor, Michigan 48109, USA.}}
\footnotetext{\textit{$^{b}$~Department of Chemical Engineering, University of Michigan, Ann Arbor, Michigan 48109, USA.}}
\footnotetext{\textit{$^{c}$~Department of Materials Science and Engineering, University of Michigan, Ann Arbor, Michigan 48109, USA.}}
\footnotetext{\textit{$^{d}$~Biointerfaces Institute, University of Michigan, Ann Arbor, Michigan 48109, USA. E-mail: sglotzer@umich.edu}}


\footnotetext{\ddag~These authors contributed equally to this work.}

\footnotetext{\textsection~Present address: Department of Cellular and Molecular Pharmacology, University of California, San Francisco, California 94158, USA.}



\section*{Introduction}

The rich self-assembly behavior of hard polyhedra has been studied in increasing detail in recent years\cite{Haji-Akbari2009,Agarwal2011,Damasceno2012a,Damasceno2012b,Henzie2012,Ni2012}.
Hard polyhedra crystallize through entropic interactions\cite{Frenkel1999,VanAnders2014a,VanAnders2014b}.
Of the many studies, most have been performed of one-component systems.
Studying the self-assembly of mixtures of polyhedra represents a natural extension to the work done on single-component systems.
Binary mixtures have many applications, \textit{e.g.}, the suitability of binary colloidal crystals for photonic applications \cite{Hynninen2007,Wan2009}.

Regular octahedra and tetrahedra with identical edge lengths tile space at a composition of 1:2, respectively.
Because both shapes have very few facets, they represent one of the simplest space-tessellating binary mixtures.
Previous studies of this system did not observe the self-assembly of the space-filling structure unless attractive interactions were added \cite{Khadilkar2012}.
Other studies on binary mixtures of different shapes produced disordered mixed lattices, but not the formation of ordered structures\cite{Khadilkar2013,Khadilkar2014}.

Experimentally, both shapes have been studied in recent years.
Octahedra were found to form a variety of structures, among them a low-density body-centered cubic (\textit{bcc}) structure with vertex-sharing octahedra\cite{Zhang2011},
vertex- and edge-sharing packings derived from the binary structure\cite{Lu2008},
a high-pressure lithium phase ($cI16$-Li \cite{Hanfland2000}), induced by a depletion effect\cite{Henzie2012},
lower-symmetry structures with stacking variants\cite{Chang2008,Xie2009,Zhang2009},
and distorted \textit{bcc} arrangements that likely correspond to the densest-packing structure\cite{Song2006,Jing2008,Henzie2012}.
Most recently, a new monoclinic assembly of octahedra was observed\cite{Gong2016}.
Tetrahedral nanoparticles have been more difficult to synthesize and experiments have not yet reproduced the predicted quasicrystal, but rather a superlattice structure\cite{Boles2014}.
Millimeter-sized frictional tetrahedra and macroscopic tetrahedral dice were found to produce only jammed \cite{Neudecker2013} and random packings\cite{Jaoshvili2010}.

To our knowledge, mixtures of octahedra and tetrahedra have not been investigated experimentally.
Superlattices of other combinations of two shapes have been studied recently, \textit{e.g.}, spheres and octahedra, as well as spheres and cubes \cite{Lu2015}.
The coexistence of multiple solid phases, on the other hand, had previously been discovered in systems of spheres with two different radii\cite{Filion2011}, but has not directly been observed in a number of studies on binary systems with polyhedra\cite{Khadilkar2012,Khadilkar2013,Khadilkar2014}.

Here we use Monte Carlo simulations to investigate the self-assembly of mixed octahedra and tetrahedra.
Other phases known to self-assemble in the one-component systems are a dodecagonal quasicrystal formed by tetrahedra \cite{Haji-Akbari2009}, as well as a distorted \textit{bcc} structure of octahedra\cite{Damasceno2012b}, both of which may compete with a co-assembled binary phase.
We investigate the phase behavior of different stoichiometries of hard tetrahedra and hard octahedra. We find both pure phases that have been reported in one-component systems coexisting with the binary crystal that is reported in self-assembly simulations here for the first time.

\section*{Methods}

We performed hard-particle Monte Carlo (MC) simulations with HPMC\cite{Anderson2013,Anderson2016}, a plugin to the HOOMD-blue software\cite{Anderson2008,HOOMD}, in the NVT thermodynamic ensemble to study the behavior of mixtures of hard tetrahedra and octahedra.
The polyhedra are modeled as perfectly faceted shapes of unit length, with sharp vertices and edges.
Simulations were carried out in cubic boxes with periodic boundary conditions, containing $10,648$ particles for each state point comprising the phase diagram.

The phase diagram was sampled at packing fractions $\phi = 0.50-0.65$ ($\Delta\phi = 0.01$)
for octahedra-tetrahedra compositions (O:T) = 0:10, 1:9, 2:8, 3:7, 4:6, 5:5, 6:4, 7:3, 8:2, 9:1, 10:0, as well as O:T = 1:999, 5:995, 1:99, 2:98, 5:95, and the special stoichiometry 1:2.
A Steinhardt order parameter was used to categorize particles as either liquid or solid, based on their local environment\cite{Steinhardt1983,Keys2011}.
This allowed for subsequent crystal structure identification and mapping of the phase diagram.

\section*{Results and discussion}

\subsection*{Phase diagram}

\begin{figure*}
\centering
\includegraphics[width=\textwidth]{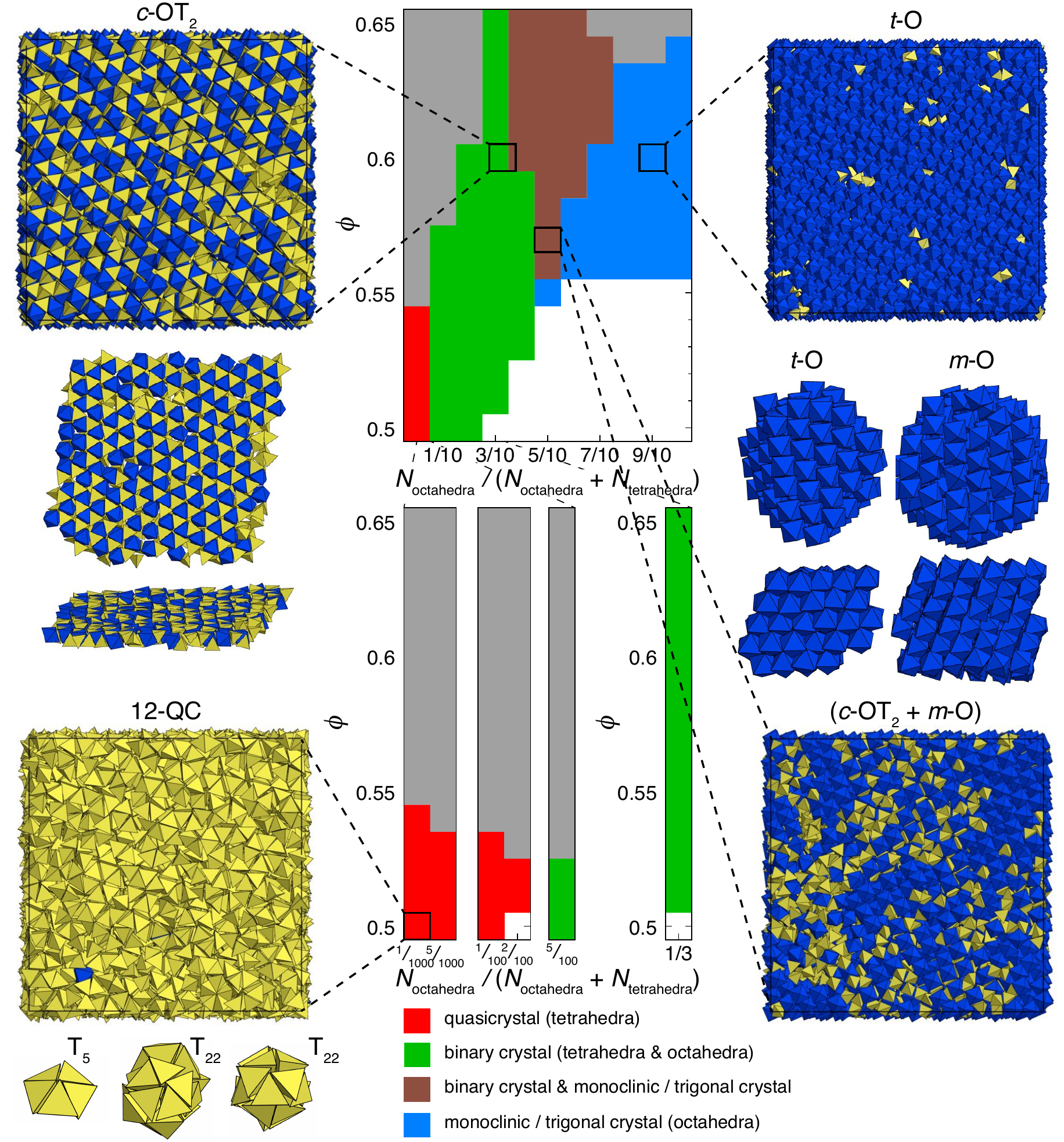}
\caption{Phase diagram of the binary system octahedra:tetrahedra displaying several crystalline phases. The entire compositional region is shown in the \textit{upper middle}. Two highlighted regions are shown \textit{below} -- the tetrahedra-rich regime and the ideal composition of the space tessellation of octahedra and tetrahedra with O:T = 1:2. Compositions are given in terms of the number of octahedra $N_\textrm{octahedra}$ divided by the total number of particles $N_\textrm{octahedra}+N_\textrm{tetrahedra}$.
Samples from self-assembly simulations are shown for all relevant phase-diagram regions (tetrahedra -- yellow; octahedra -- blue): the dodecagonal quasicrystal (12-QC) comprised of tetrahedra \textit{(lower left)}, the binary cubic phase (\textit{c}-OT$_2$) \textit{(upper left)}, the trigonal phase comprised of octahedra (\textit{t}-O) \textit{(upper right)}, and a sample containing coexisting \textit{c}-OT$_2$ and \textit{t}-O regions \textit{(lower right)}.
Structural motifs of the 12-QC are shown (\textit{left to right:} pentagonal bipyramid T$_5$; topview, and sideview of the T$_{22}$ building block), on the \textit{middle left}, a representative layer that builds the \textit{c}-OT$_2$ structure, and on the \textit{middle right} a topview and sideview of the layers of octahedra in the \textit{t}-O structure.
Gray regions in the phase diagram signify simulations that did not fully crystallize
within the performed number of MC sweeps.
White regions did not exhibit any ordered phases and can be regarded as fluid.}
\label{phasediagram_structures}
\end{figure*}

Three crystal structures were observed for the stoichiometries and packing fractions studied, with two solid-solid coexistence regions between them. A schematic phase diagram is given in Fig.~\ref{phasediagram_structures}.
The quasicrystal 12-QC is shown in red, the pure binary crystal phase OT$_2$ is shown in green, the pure crystals of octahedra are shown in blue, and the coexistence of binary and pure octahedral crystals is shown in brown.
Snapshots of all crystal structures are depicted as well.

At low numbers of octahedra and around 50\% packing fraction ($\phi = 0.50$), the tetrahedra self-assemble into the dodecagonal quasicrystal (12-QC) that had been reported for systems composed solely of tetrahedra\cite{Haji-Akbari2009}.
As the octahedra content increases, the quasicrystal phase quickly disappears:
it is observed up to stoichiometries O:T = 0.01:0.99, but not in simulations with O:T =  0.1:0.9.
The packing fractions at which the dodecagonal quasicrystal forms are at the low end of the investigated phase diagram, at packing fractions of 50--53\%.

In the low O:T regions of the phase diagram, the tetrahedra begin forming the 12-QC within about 8 million MC sweeps.
Together with the octahedra, they are expected to form a more stable phase, the space-filling OT$_2$ structures, but the octahedra diffuse relatively slowly through the dense quasicrystal. 
For O:T = 0.02:0.98 at $\phi$ = 52\%, for example, there are no hints of order in the arrangement of octahedra until after nearly 160 million MC sweeps.
Thus, a rough sweep of the phase diagram does not detect a coexistence region between the 12-QC and OT$_2$ phases, however, extensive runs in the intermediate compositional region suggest that the binary crystal will form whenever octahedra are introduced to the quasicrystal.
Additional simulations at $\phi$ = 52\% revealed
local motifs of the binary structure
at O:T =  0.02:0.98 and 0.03:0.97 after 140 million MC sweeps, whereas the addition of octahedra to a system consisting mostly of tetrahedra further suppresses the growth of the 12-QC at O:T = 0.04:0.96
and no 12-QC phase was observed at 140 million MC sweeps.

\begin{figure}[h!]
\centering
\includegraphics[width=0.45\textwidth]{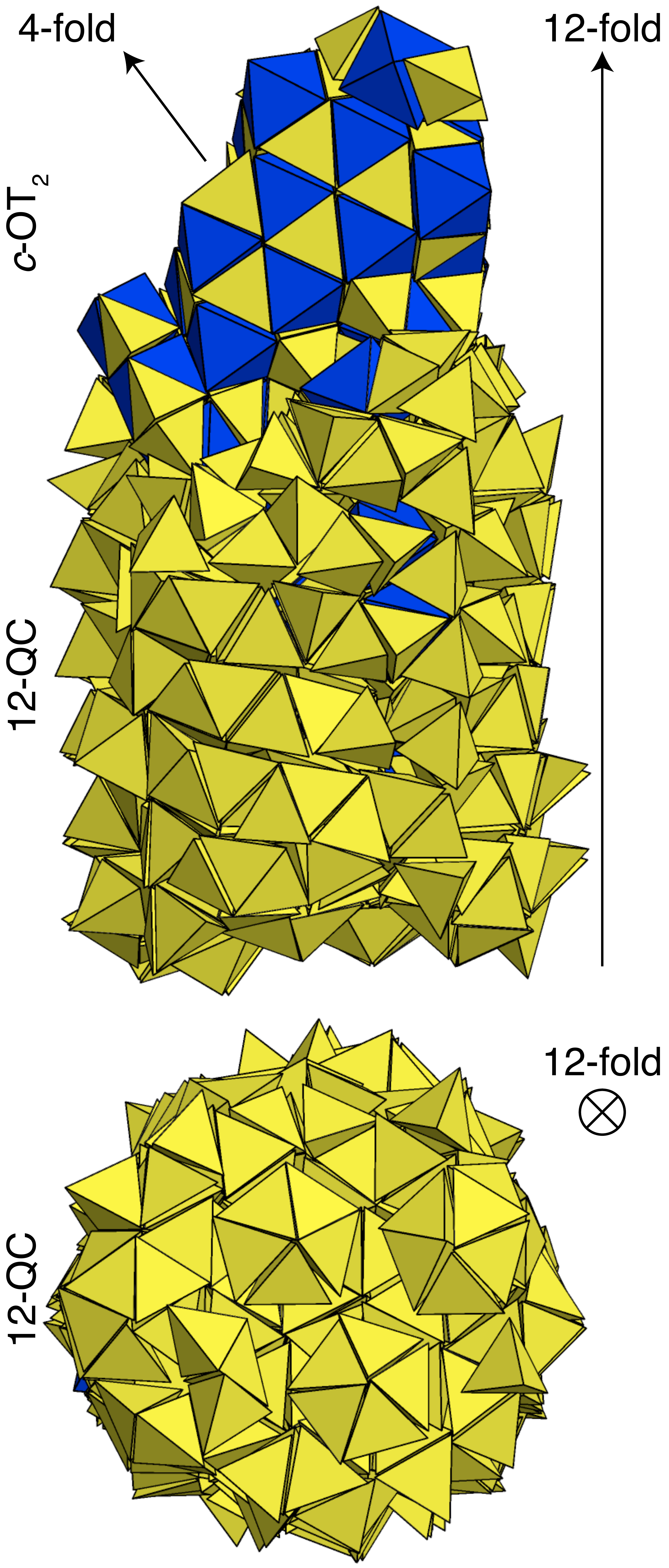}
\caption{Oriented intergrowth of the 12-QC structure and the binary phase. The 12-fold symmetry axis of the quasicrystal is oriented in the vertical direction and the intergrowth angle becomes visible in the orientation of the cubic binary phase.}
\label{QCbinIntergrowth}
\end{figure}

When octahedra are present next to the quasicrystal, they begin to aggregate in binary face-to-face alignments with tetrahedra after approximately 40 million MC sweeps.
As the mixture begins to form the binary crystal, octahedra take the place of a dimer of two tetrahedra along one of the columns of the quasicrystal.
Correspondingly, the angle at which the binary crystal forms is offset from the main axis of the quasicrystal, as shown in Fig.~\ref{QCbinIntergrowth}.
Any octahedra introduced into the quasicrystal eventually form the \textit{c}-OT$_2$ structure with the corresponding amount of tetrahedra.

At intermediate compositions of octahedra and tetrahedra, the binary OT$_2$ structure
forms, with a small amount of stacking faults.
The binary crystal forms with ease at packing fractions 54--61\% and O:T = 3:7--6:4 around the stoichiometry of the ideal structure, O:T = 1:2. The high flexibility of the packing with respect to the amount of tetrahedra that are available in the polyhedral mixture likely stems from the ability of the structure to accommodate a number of tetrahedral voids.

At high O:T ratios, the system forms pure packings of octahedra. They can mostly be observed for O:T = 7:3--9:1 and at packing fractions of 56--60\%. A large coexistence region is located between the stability regions of the binary crystal and the crystal of octahedra, where both ordered phases were detected.

\subsection*{Structures and motifs}

\begin{figure}[h!]
\centering
\includegraphics[width=0.5\textwidth]{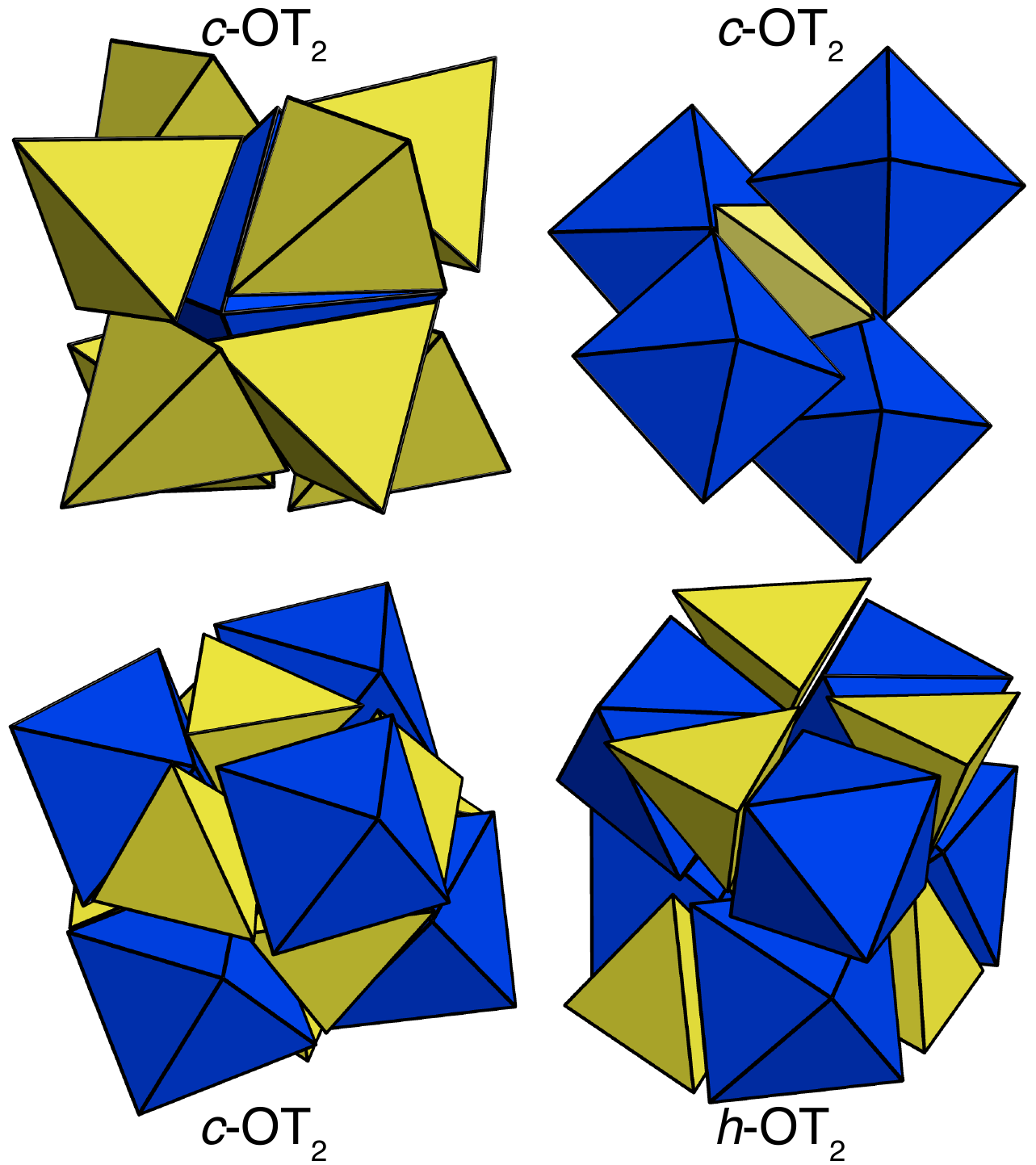}
\caption{Structural motifs observed in binary assemblies of octahedra and tetrahedra. Tetrahedra are depicted in yellow, octahedra in blue color.
Shown are: one octahedron surrounded by 8 tetrahedra as found in $c$-OT$_2$ \textit{(upper left)}, one tetrahedron surrounded by four octahedra as found in $c$-OT$_2$ \textit{(upper right)}, $c$-OT$_2$-type stacking with alternating octahedra and tetrahedra in the stacking direction \textit{(lower left)}, and $h$-OT$_2$-type stacking with octahedra stacked on top of octahedra and tetrahedra on tetrahedra \textit{(lower right)}.
}
\label{localmotifs}
\end{figure}

The most important motifs that occur in the crystal structures found in this system are depicted in Figs.~\ref{phasediagram_structures} and \ref{localmotifs}.

Tetrahedra alone form the same arrangements that were previously observed in the dodecagonal quasicrystal\cite{Haji-Akbari2009} (12-QC): columns of pentagonal bipyramids alternating with rings of six dimers that are arranged in a dodecagonal square-triangle tiling.
This local arrangement of 22 tetrahedra has also been found to occur in spherical confinement\cite{Teich2016}.

In the binary mixture, octahedra and tetrahedra are arranged alternatingly and face-to-face in a dense layer. Within the layer, each triangular face is shared between an octahedron and a tetrahedron.
These binary octahedra-tetrahedra (O:T = 1:2) layers can be stacked in two different ways, while having polyhedra between layers align face-to-face: 
alternatingly -- with only O--T contacts between layers -- or vice versa, with only O--O and T--T contacts.
Employing only one kind of stacking results in two regular honeycombs, \textit{i.e.}, space-filling tessellations: the alternated cubic honeycomb and the gyrated honeycomb, respectively.
Both structures have the same composition as each individual layer (O:T = 1:2; OT$_2$) and can be combined in infinitely many different stacking variants.
We label the basic cubic and hexagonal structures by their symmetries: \textit{c}-OT$_2$ and \textit{h}-OT$_2$.
The arrangements of octahedra and tetrahedra correspond to the layout of octahedral and tetraheral interstices in the cubic and hexagonal close-packed structures, \textit{ccp} and \textit{hcp}.

Octahedra alone have been found to pack in two different phases, a monoclinic\cite{Gong2016} and a trigonal one, \textit{m}-O and \textit{t}-O, the latter being the conjectured densest packing also known as the Minkowski phase\cite{Minkowski1904,Betke2000}, which has previously been confirmed by packing simulations\cite{Torquato2009}.
There are no shear planes in the \textit{t}-O phase, which entropically favors the \textit{m}-O structure, except at the highest packing fractions.
In most mixed systems of octahedra and tetrahedra, both phases occur; an increased content of tetrahedra seems to facilitate the formation of the \textit{m}-O structure, while the pure system of octahedra exhibits mainly \textit{t}-O.
The fact that octahedra self assemble at larger packing fractions than tetrahedra is probably due to the circumstance that neither of the two octahedra packings has well-aligned face-to-face contacts. Instead, two neighboring polyhedra are always shifted with respect to one another, while a large portion of tetrahedra within the 12-QC structure exhibit face-to-face contacts that are much better aligned.

\section*{Summary \& Conclusions}

We have shown for the first time that binary crystal structures self-assemble from mixtures of hard, non-interacting polyhedra.
We observed that this binary phase coexists with both neighboring phases in self-assembly simulations: one comprised of only tetrahedra and the other of only octahedra.

In the binary phase diagram of mixtures of tetrahedra and octahedra,
we reported crystal structures self-assembling in three regions:
 the known dodecagonal quasicrystal formed by tetrahedra, assemblies of octahedra only, and the binary OT$_2$ phase.

The phase diagram contains only a very narrow coexistence region between the 12-QC and \textit{c}-OT$_2$ phases.
Generally, the quasicrystal forms at much lower packing fractions than the binary crystal.
However, even at low packing fractions, this small number of octahedra
rearrange in OT$_2$-like building units and incorporate the respective amount of tetrahedra to form the binary crystal.

There is a distinct range of stoichiometries that allows for coexistence of the OT$_2$ structure and a pure octahedral crystal. Both occur at similar packing fractions for stoichiometries around O:T = 6:4--7:3. Thus in the respective simulation runs, both crystalline phases were detected.

All in all, the phase diagram of mixtures of octahedra and tetrahedra exhibits typical features, such as pure and mixed phases, coexistence regions, \textit{etc.} The phase behavior of this entropically stabilized system thus does not fundamentally differ from that observed in systems with enthalpic interactions.

\section*{Acknowledgements}

We thank Erin G.\ Teich for helping with the analysis of polycrystalline structures within the simulation data.
This material is based upon work supported part by the U.S.\ Army Research Office under Grant Award No.\ W911NF-10-1-0518 and also by the National Science Foundation, Division of Materials Research Award \# DMR 1120923.
J.\ D.\ acknowledges support through the Early Postdoc.Mobility Fellowship from the Swiss National Science Foundation, grant number P2EZP2\_152128.
P.\ F.\ D.\ acknowledges support from the University of Michigan Rackham Pre-doctoral Fellowship Program.
This research used the Extreme Science and Engineering Discovery Environment (XSEDE), which is supported by National Science Foundation grant number ACI-1053575 (XSEDE award DMR 140129), and local computational resources supported by Advanced Research Computing at the University of Michigan, Ann Arbor.






\bibliography{BinaryPaperRefs}
\bibliographystyle{rsc} 

\end{document}